%% file: altern.tex
\documentclass[runningheads]{llncs}
\usepackage{amssymb}
\usepackage{amsmath}
\usepackage{graphicx}
\usepackage{epsfig}
\usepackage[all]{xy}
\usepackage{url}
\usepackage{texdraw}
\usepackage{float}
\usepackage{ifthen}
\usepackage[PostScript]{diagrams}

\newboolean{extabs}
\setboolean{extabs}{true}

\input{macros.tex}

\ifthenelse{\boolean{extabs}}{
\input{lncsnotes.tex}
}{
\input{rschnotes.tex}
}

\input{double.tex}
\input{triple.tex}

\newarrow{Dash} .....

\ifthenelse{\boolean{extabs}}{
\title{Alternating Product Ciphers: A Case for Provable Security Comparisons}
\subtitle{(extended abstract\footnote{to appear in \emph{Indocrypt 2013}.})}
}{
\title{Alternating Product Ciphers: A Case for Provable Security Comparisons}
}

\author{John O. Pliam\thanks{The author's position is supported in part by the U.S. National Science Foundation (NFS) Data-Scope grant CISE/ACI 1040114.}}
\institute{Institute for Data Intensive Engineering and Science (IDIES), Johns Hopkins University, 3400 N.\ Charles St., Baltimore, MD 21218, USA \email{john.pliam@jhu.edu}}

\begin{document}
\maketitle
\begin{abstract}
We formally study iterated block ciphers that alternate between two sequences of independent and 
identically distributed (i.i.d.) rounds.  It is demonstrated that, in some cases the effect of 
alternating increases security, while in other cases the effect may strictly \emph{decrease} 
security relative to the corresponding product of one of its component sequences.  
As this would appear to contradict conventional wisdom based on the ideal cipher approximation, 
we introduce new machinery for provable security comparisons.  The comparisons made here
simultaneously establish a coherent ordering of security metrics ranging from key-recovery cost 
to computational indistinguishability.

\vspace*{1em}
\end{abstract}

%
%
\notesect{Introduction}

\begin{note}{Overview}
For many decades, various issues related to product ciphers have been raised and addressed. 
A large part of Shannon's seminal work \cite{shan} is devoted to both theoretical and practical 
aspects of products, and his invocation of the pastry dough mixing analogy \cite[p.\ 712]{shan}
captures a very intuitive idea that by alternating between two weakly mixing operations, 
we should eventually achieve strong mixing.  Even today, many modern block cipher designs retain 
an element of this structure (see, e.g.\ \cite{knro}).

To model such mixing, we formalize the notion of an \emph{alternating product cipher} as an 
interleaving product of independent ciphers as depicted in Fig.~\ref{fig:altern}.  We then 
ask: \textit{How well does the mixing work, and how might it fail?}  Various outcomes seem possible.  
We present a threefold alternating product with good security expansion.  However, a related 
construction demonstrates somewhat surprisingly, an alternating product which is strictly less 
secure than the two-term product of one of the component sequences by itself.  On its face, this 
would appear to contradict an emerging conventional wisdom about multiple encryption based on the 
ideal cipher approximation, ``that double encryption improves the security only marginally [...] 
triple encryption is significantly more secure than single and double encryption'' \cite{gzmr}.
The situation demands that we explore the problem of provable security \emph{comparisons}.  We 
find that certain security orderings transcend the (somewhat artificial) boundary between 
classical and modern cryptography. 

We conclude that alternating product ciphers are, at a fundamental level, different from two-term 
products.  Ascertaining their security is more nuanced and they provide evidence of further limits 
on the applicability of the ideal cipher approximation (see also \cite{blac}).
\end{note}

\begin{note}{Motivation}
We are initially motivated by how we might generalize the question, ``is a cipher a group?'', in the 
case of alternating products.  Roughly, an encryption function $E:K\times M\map M$ is said to have the 
\emph{group property} \cite[p.673]{shan}, if for each key pair $(k_1,k_2)$, there is another key 
$k\in K$ such that $E(k_2, E(k_1, p))=E(k,p)$ for each plaintext $p\in M$.  Equivalently, the product 
of the cipher with itself produces \emph{no new} permutations.

The group property obviously affords the cryptanalyst considerable advantage, if only because it 
reduces the cost of brute-force search against the product.  Understandably, the question was raised 
as a possible weakness to multiple encryption schemes of DES \cite{krs}.  These concerns were 
dismissed with increasing strength as researchers showed that DES was not likely to be a group 
\cite{krs}, that it is not a group \cite{cmwn} and that it generates a large group \cite{wern}.

Questions about whether a cipher is a group or whether multiple encryption improves security are 
really questions about ordering.  That is to say, rather than quantifying specific 
models of attack against fixed encryption systems, we seek to establish the \emph{correct ordering} 
between constructs of interest.  In the case of alternating products, we find the comparison between 
$XYZ$ and $XZ$ to be the most intriguing since our intuition suggests that inserting statistically 
independent $Y$ in between $X$ and $Z$ should improve security.  Thus in comparing the
two products of Fig.~\ref{fig:cmp}, we find that the order itself depends on the internal 
structure of the constituents.
 
\begin{figure}[ht]
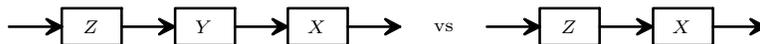

{\scriptsize
$$
\begin{array}{rcccl}
{\triple{Z}{Y}{X}} && \;\;\;\mathrm{vs}\;\;\; && \double{Z}{X}
\end{array}
$$
}
\caption{Motivating comparison between alternating product $XYZ$ and $XZ$.}
\label{fig:cmp}
\end{figure}
\end{note}

\begin{note}{Toward Coherent Security Ordering}
Motivated by the above, we start by quantifying how the permutation count of an alternating product 
can grow or shrink.  Numerically comparing these counts offers one possible ordering, since integers 
are \emph{totally ordered} (every pair is comparable).  But we argue that by relaxing this notion and 
considering \emph{partial/pre orders} on ciphers, we pave the way to stronger and more broadly 
applicable security comparisons.  There is a lucrative trade-off here: if we give up comparing 
\emph{every} pair of ciphers, we are left with a \emph{more meaningful} ordering of the remainder.

One powerful order (known to not be total) is \emph{majorization}, and a great many interesting 
real-valued security metrics are known to respect majorization.  These are called \emph{Schur-convex} 
(concave if they reverse it).  This covers the case of zero data complexity in the far left of
Fig.~\ref{fig:schurfkt}; if a majorization relationship between two ciphers can be established, then 
the ciphers are also ordered by the real values of \emph{any} Schur-convex(concave) function.

Better still, the comparisons of Fig.~\ref{fig:cmp} in this paper possess additional structure,
facilitating a coherent ordering of security metrics in arbitrary data complexity $q$.  Specifically, 
we show in Sect.~\ref{sect:modscmp}, that nonadaptive chosen-plaintext attack (ncpa) advantage 
\cite{mrs,dcor} as well as conditional guesswork \cite{cis} are such metrics.  This is depicted in 
the two rightmost diagrams in Fig.~\ref{fig:schurfkt}.
\begin{figure}[ht]
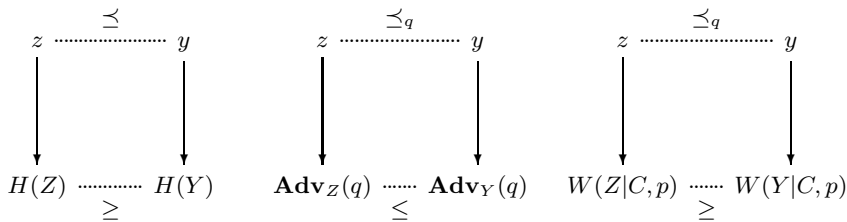

$$
\begin{array}{ccccc} 
\begin{diagram}
z & \rDash^\majby & y  \\
\dTo & & \dTo \\
H(Z) & \rDash_\geq & H(Y)
\end{diagram}
& \;\;\;\;\;\;\;\;\;\; &
\begin{diagram}
z & \rDash^\qmajby & y  \\
\dTo & & \dTo \\
\adv_Z(q) & \rDash_\leq & \adv_Y(q)
\end{diagram}
& \;\;\;\;\;\;\;\;\;\; &
\begin{diagram}
z & \rDash^\qmajby & y  \\
\dTo & & \dTo \\
W(Z|C,p) & \rDash_\geq & W(Y|C,p)
\end{diagram}
\end{array}
$$
\caption{We establish a coherent ordering of the ciphers in Fig.~\ref{fig:cmp} by showing the 
consistency of a broad range of security metrics, crossing the divide between information-theoretic 
and modern cryptography.}
\label{fig:schurfkt}
\end{figure}
\end{note}

%
%
\notesect{Preliminaries}

\begin{note}{Prerequisites}
We've tried to make this paper readable by nonspecialists conversant in contemporary cryptography.  
But in order to follow the proofs, we assume additional familiarity with the basics of permutation 
groups, probability theory and representation theory, referring the reader to \cite{rotm}, \cite{diac}.
We also exploit aspects of the theory of majorization, treated very well in \cite{moa}.  Since that is 
critical here, we provide a brief summary in \S\ref{n:maj}.
\end{note}

\ifthenelse{\boolean{extabs}}{}
{
\input{grpact}

\input{doubcos}
}

\begin{note}{Shannon's Model}
\label{n:shamo}
We formalize Shannon's model \cite{shan} by representing an 
encryption system as a permutation-valued random variable.  Precisely, 
given message space $\msg$, let $G$ be some subgroup of the full symmetric group $\sym{\msg}$ 
of all permutations on $\msg$.  An encryption system on $\msg$, or $G$-cipher for short, is a
 $G$-valued random variable $X$.  As a notational convention, for a $G$-cipher $X$ (always uppercase), 
we shall denote the probability distribution from which it is drawn by lowercase function 
$x:G\map\reals$ and write $X\sim x$.  

Shannon observed that the set of encryption systems is endowed with the structure of an 
\emph{unital associative algebra}\footnote{These days, it would be identified as the \emph{group 
algebra} $\reals G$.} whose \emph{sum} and \emph{product} correspond to parallel and series 
composition, respectively.  The composition in series of independent $G$-ciphers $X,Y$ gives the 
notion of a product cipher $Z=XY$, which survives to this day.  It is a standard observation
that the probability distribution of the product $z(g)=\prob{XY=g}$, is given by the 
\emph{convolution} $z=x*y$:
\begin{equation}
\label{eqn:conv}
z(g) = \sum_{h\in G} x(gh^{-1})y(h).
\end{equation}
\end{note}

\notesect{Models For Security Comparison}
\label{sect:modscmp}

After a brief review of the theory of majorization, we explore ways in which 
claims of security ordering may be rigorously established as in Fig.~\ref{fig:schurfkt}.

\begin{note}{Majorization  and Schur-Convexity} 
\label{n:maj}
Given vectors $x,y\in\rpn$ we say $x$ is \textit{majorized} by $y$ and write $x\majby y$, 
if their $l_1$ norms agree and for each $1\leq k\leq n$,
$$
\sum_{i=1}^k x_{[i]} \leq \sum_{i=1}^k y_{[i]},
$$
with the values rearranged by $x_{[1]} \geq x_{[2]} \geq \cdots \geq x_{[n]}$ and similarly 
for $y$. The vector $x_\desc$ is the decreasing rearrangement of $x$ (so $x_{[i]} = (x_\desc )_i$).
Majorization is a \textit{preorder} relation, so not all vectors are comparable in this 
way.  We have, from the \emph{\hlp\ theorem}, that $x\majby y$ is equivalent to the existence of a 
\textit{doubly-stochastic} matrix $D$ such that $x=Dy$.  Furthermore, by the \emph{\bvn\ theorem},
such a matrix is a convex sum of permutations, so $x\majby y$ means:
\begin{equation}
\label{eqn:hlpbvn}
x=\sum_{\pi\in\sym{n}} p_\pi \pi\cdot y.
\end{equation}
For 
our purposes, the vectors will usually be probability distributions, each with $\l_1$ norm of $1$.
It is readily verified that the uniform distribution $u=(1/n,\ldots,1/n)$ has $u\majby x$ for all
probability vectors $x$.  If $x\majby y$ and $y\majby x$, then $x$ is a permutation of $y$.  If 
$x\majby y$ but $x$ is \emph{not} a permutation of $y$, we'll write $x\smajby y$. 

Certain useful real-valued functions respect or reverse majorization.  So if $\phi : \rpn\map\reals$  
has $\phi(x)\leq\phi(y)$ ($\phi(x)\geq\phi(y)$) whenever $x\majby y$, we call $\phi$ 
\textit{Schur-convex (concave)}.  If a Schur-convex (concave) function additionally satisfies 
$\phi(x) < \phi(y)$ ($\phi(x) > \phi(y)$) when $x\smajby y$, we call $\phi$ \textit{strictly 
Schur-convex (concave)}.

Examples and applications abound throughout science and engineering (see e.g. \cite{ciug} for
an interesting information-theoretic treatment).  In particular, 
\emph{Shannon entropy}, \emph{R\'{e}nyi entropy} and \emph{guesswork} \cite{mass} are strictly 
Schur-concave.  Furthermore \emph{marginal guesswork} \cite{indo} and Bonneau's 
\emph{$\alpha$-guesswork}
\cite{bonnphd} are Schur-concave.  For further details, see \cite[pp.\ 562--564]{moa} and 
\ifthenelse{\boolean{extabs}}{\cite[Appx.]{ic13full}}{Appx.~\ref{n:agw}}.  As remarked, 
majorization treats the case of zero data complexity, which is sometimes useful by itself.
\end{note}

\begin{note}{Nontrivial Data Complexity}
For an adversary with access to $q$ plaintext-ciphertext pairs or equivalently $q$ queries to a 
chosen-plaintext oracle, we can often identify a vector mapping $\sigma:V\map\hat{V}$ and
a Schur-convex function $\phi_q$ on $\hat{V}$ measuring in some way the cipher's resistance to attack.
If $z\majby y$ in $V$ has additional structure so that $\hat{z}\majby\hat{y}$ in $\hat{V}$,  we write 
$z\majby_{\;\phi_q\circ\sigma} y$ or just $z\qmajby y$ when clear from context.  This situation affords 
meaningful security comparisons for arbitrary data complexity.  A proof of the following is 
\ifthenelse{\boolean{extabs}}{sketched in the appendix and proved in the full version 
\cite{ic13full}.}{proved in the appendix.}

\begin{thm}
\label{thm:qschur}
Given data complexity limit $q$ and $G$-ciphers $X\sim x$, $Y\sim y$ and $Z\sim z$ with $Z=XY$,
we have (for appropriate choices of $\sigma$)
\begin{enumerate}
\item $z\qmajby y$ for conditional guesswork: $W(E|C,p)$, $p\in\msg^{(q)}$ is Schur-concave as a 
function of $\hat{e}$, 
\item $z\qmajby y$ for distinguishing advantage: $\ncpa{E}(q)$ is Schur-convex as a function of 
$\hat{e}$,
\end{enumerate}
Here $E\sim e$ is a generic argument.
\end{thm}
The relationship $z\qmajby y$ can arise in many ways, but for our purposes, we'll use the fact that 
$Z=XY$.
\end{note}

\notesect{Alternating Product Ciphers}

\begin{note}{The Formal Definition}
We may now give a formal definition of an alternating product followed by an example.
\begin{defn} 
\label{defn:ap}
An \textbf{alternating product} is the product of independent $G$-ciphers alternating between two 
sequences of i.i.d.\ $G$-ciphers.
\end{defn}

\begin{ex} 
\label{ex:altern}
Let $\{X_i\}_{i=1}^{r+1}$ be i.i.d.\ $G$-ciphers and let $\{Y_i\}_{i=1}^r$ be distinct i.i.d.\ 
$G$-ciphers.  Then $E=X_{r+1}Y_rX_r\cdots Y_1X_1$ is an alternating product of $X_i$ and $Y_i$.  
Notice that Def.\ \ref{defn:ap} permits either an even or an odd number of components in the 
product.  We can imagine $E$ as alternating between the ``factors'' of two products $X=X_{r+1}
\cdots X_1$ and $Y=Y_r\cdots Y_1$ as depicted in Fig.\ \ref{fig:altern} below.
\end{ex}

\begin{figure}[ht]
{\scriptsize
$$
\input{altprod.tex}
$$
}
\caption{An alternating product cipher seen as an interleaving of the terms of two iterated 
ciphers.}
\label{fig:altern}
\end{figure}
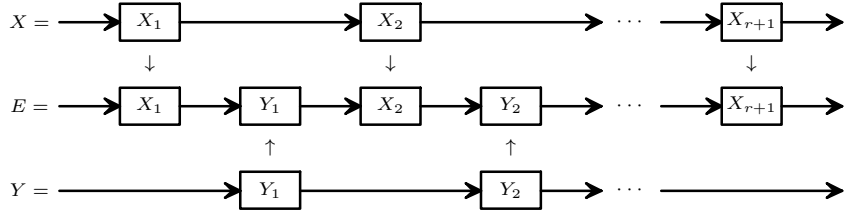
\end{note}

\begin{note}{Threefold Mixing Convolutions and Double Cosets}
While most of this paper is devoted to alternating products, we treat a slightly more general
case in this section to explicate a key observation, namely how mixing in typical iterated block 
ciphers is related to expansion along double cosets when randomness enters via a subgroup operation 
(like the XOR-ing of round subkeys).

Consider a threefold product $T=XYZ$ with $Z$ confined to subgroup $K\leq G$, $X$ confined to 
subgroup $H\leq G$, and $Y$ deterministically taking single value $\pi\in G$.  To understand how 
the convolution $t=x*y*z$ decomposes, it's instructive to employ an action of $g\in G$ on 
functions $\phi:G \map \reals$ taking $\phi \mapsto \phi\circ g^{-1}$, in other words
$(g\cdot \phi)(f) = \phi(g^{-1}f)$.  With this in mind, we have the following useful lemma.
\end{note}

\begin{lem}
\label{lem:confine}
If the support of $\phi$ is confined to a left coset $kH$, then the support of $g\cdot\phi$ is 
confined to $gkH$.
\end{lem}
\begin{pf} Assume $\supp(\phi) \subseteq kH$, and observe that
$f \in \supp(g\cdot\phi) 
  \implies (g\cdot\phi)(f) \neq 0 
  \implies \phi(g^{-1}f) \neq 0 
  \implies g^{-1}f \in \supp(\phi) 
  \implies g^{-1}f \in kH
  \implies f \in gkH$.
Thus $\supp(g\cdot\phi) \subseteq gkH$.
\end{pf}

\vspace*{1em}

It should be intuitively clear that $T$ is spread out over the double coset $H\pi K$, but the
following stronger result details some of the mechanics of the mixing, facilitating deeper 
security comparisons.

\begin{thm}
\label{thm:3conv2cos}
The cipher $T$ has $\supp(t)\subseteq H\pi K$ and its distribution is a convex direct sum
$$
t = \bigoplus_{i=1}^m \alpha_i z_i,
$$
of $m=\idx{H}{H\cap\lconj{\pi}{K}}$ distinct probability vectors.  Furthermore, each $z_i$ is 
majorized by $z$.
\end{thm}

\begin{pf}
The three terms in $T=XYZ$ are given by probability distributions $x(g), y(g)$ and $z(g)$ with 
$\supp(x)\subseteq H$, $\supp(y)=\{\pi\}$ and $\supp(z)\subseteq K$.
First note that by the 
associativity of product ciphers, we may write $T=X(YZ)$, and the inner convolution $z'=y*z$ yields,
$$
z'(g) = \sum_{h\in G} y(h)z(h^{-1}g) = z(\pi^{-1}g) = \pi \cdot z(g),
$$
which by Lem.~\ref{lem:confine}, precisely describes a function confined to $\pi K$.

Now the outer convolution $t=x*z'$ yields
$$
t(g) = \sum_{h\in G} x(h)z'(h^{-1}g) = \sum_{h\in H} x(h)z'(h^{-1}g) = \sum_{h\in H} x(h) h\cdot z'(g).
$$
Recognizing that $z'$ is confined to $\pi K$, it is natural according to Lem.~\ref{lem:confine} to 
collect terms for which $h\cdot z'$ is confined to the same left coset of $K$. Indeed, recall
that the left action of $H$ on left cosets $G/K$ (the supports of the various 
$h\cdot z'$) is equivalent to the double coset action $H\backslash G/K$.   We may decompose the 
orbit $H\pi K = \bigcup_{i=1}^m \lambda_i K$, where the orbit size $m=\idx{H}{H\cap\lconj{\pi}{K}}$ 
is given by the orbit-stabilizer theorem with
$$
S\defeq\stab_{H\backslash(G/K)}(\pi K)= H\cap \lconj{\pi}{K}.
$$
Furthermore, we recognize $\setst{h\in H}{h (\pi K) = \lambda_i K} = h_i S$, for some left 
transversal $\{h_i\}$ of $S$ in $H$.  This gives us a recipe for collecting terms,
\begin{eqnarray*}
t = \sum_{h\in H} x(h) h\cdot z'
  = \sum_{i=1}^m \sum_{h\in h_iS} x(h) h\cdot z'
  &=& \sum_{i=1}^m  x(h_iS) 
    \underbrace{\left( \frac{1}{x(h_iS)} \sum_{h\in h_iS} x(h) h\cdot z'\right)}_{\defeq\;z_i} \\
  &=& \sum_{i=1}^m  x(h_iS) z_i,
\end{eqnarray*}
where, by construction, each $z_i$ is confined to left coset $\lambda_i K$, so the sum is a direct 
sum.  By the \hlp\ and \bvn\ theorems,
$$ 
z_i = \frac{1}{x(h_iS)} \sum_{h\in h_iS} x(h) h\cdot z'
$$
is a convex sum of permuted copies of $z'$, assuring majorization $z_i\majby z'\majby z$.  Finally,
taking $\alpha_i=x(h_iS)$ yields $\sum_i\alpha_i=1$ and the theorem is proved.
\end{pf}

\vspace*{1em}
Uniform distributions simplify the matter. 

\begin{coro}
\label{coro:unif2cos}
When both $X$ and $Z$ are uniformly distributed, $T$ is uniformly distributed on $H\pi K$.
\end{coro}

\begin{pf}
The uniformity of $Z$ implies $z\majby z_i\majby z$, so each $z_i$ is uniform on $\lambda_i K$. The 
uniformity of $X$ implies
$$
t = \sum_{i=1}^m x(h_iS) z_i = \frac{|S|}{|H|} \sum_{i=1}^m z_i
  = \frac{|H\cap\lconj{\pi}{K}|}{|H|} \sum_{i=1}^m z_i = \frac{1}{m} \sum_{i=1}^m z_i,
$$
which precisely describes a function uniform on $H\pi K$.
\end{pf}

\vspace*{1em}
The following is immediate.

\begin{coro}
\label{coro:smaj}
If in addition to the conditions of the previous corollary, $|H\pi K|>|K|$ then $t\smajby z$.
\end{coro}

\begin{rem} 
For noncommutative $G$, the additional condition in Coro.~\ref{coro:smaj} is actually the typical 
case, even when $H=K$.  For the remainder of this paper we assume $H\neq\lconj{\pi}{H}$ 
so that $\idx{H}{H\cap\lconj{\pi}{H}}>1$ and thus $|H\pi H|>|H|$.  This is always true in a simple 
group since $H=\lconj{\pi}{H}$ means $H$ is normal.
\end{rem}

\notesect{Applications}

\begin{note}{An Expanding Alternating Product} 
If, in Coro.~\ref{coro:smaj}, we further impose $H=K$, we obtain an alternating product cipher in 
the sense of Def.~\ref{defn:ap}.

\begin{prop}
\label{prop:expand}
The alternating product cipher $T=XYZ$ is more secure than $D=XZ$ when $X$ and $Z$ are uniform 
$H$-ciphers and $Y$ is deterministic on $\{\pi\}$, in the sense that:
\begin{enumerate}
\item[\emph{a).}] The mixing of permutations in $T$ produces dramatically more than $D$.
\item[\emph{b).}] $t\smajby d$, so by \emph{any} strictly Schur-convave security metric, $T$ is more secure than $D$.
\item[\emph{c).}] $t\qmajby d$, so by the security metrics of Thm.~\ref{thm:qschur}, $T$ is no less secure than $D$.
\end{enumerate}
\end{prop}

\begin{pf}
For (a) and (b), we need only apply Coro.~\ref{coro:unif2cos} and  Coro.~\ref{coro:smaj}. For (c),
observe that since $D=XZ=Z$, $T=XYZ=(XY)D$, we have the necessary product relation for 
Thm.~\ref{thm:qschur}.
\end{pf}
\end{note}

\begin{note}{A Collapsing Alternating Product}
\label{n:countereg}
Now let $H$ be a subgroup of $G$ and let $\pi\in G$ fall strictly outside $H$ (so $H\neq \pi H$).  
Consider three independent $G$-ciphers $X,Y,Z$, where both $X$ and $Z$ are uniformly distributed 
on the left coset $\pi H$ and $Y$ takes the value $\pi^{-1}$ deterministically.   We seek to 
compare the products $T=XYZ$ vs $D=XZ$.  Note that since $X$ and $Z$ are i.i.d., and $Y$ is 
independent of these, $T$ is also an alternating product.  We have the following.

\begin{prop}
\label{prop:countereg}
The product cipher $D$ is more secure than the alternating product $T$ in that:
\begin{enumerate}
\item[\emph{a).}] The mixing of permutations in $D$ produces dramatically more than $T$.
\item[\emph{b).}] $d\smajby t$, so by \emph{any} strictly Schur-convex security metric, $D$ is more secure than $T$.
\item[\emph{c).}] $d\qmajby t$, so by the security metrics of Thm.~\ref{thm:qschur}, $D$ is no less secure than $T$.
\end{enumerate}
\end{prop}

\begin{pf}
Without loss of generality, we may drop the trailing $\pi$, which poses no cryptanalytic barrier.  We
compare instead the products $T'=X'YZ$ and $D'=X'Z$, with $X'$ uniform on $H$.
Writing $T'=X'(YZ)$ the inner convolution $v=y*z$ trivially reduces to uniform on $H$. In this way, 
$T'=X'V$ is the double encryption of Prop.~\ref{prop:expand}.  On the other hand, $D'$ is uniform 
on $H\pi H$ since it is of the form of the triple product of Prop.~\ref{prop:expand}.  The desired 
result follows at once from this reversal of roles of the double and triple product from 
Prop.~\ref{prop:expand}.
\end{pf}
\end{note}

\begin{note}{A Collapsing General Alternating Product}
Consider again the general alternating product cipher of Fig.~\ref{fig:altern} and 
Ex.~\ref{ex:altern}, only now with each $X_i$ uniform on $\pi H$ and each $Y_i$ 
deterministically taking $\pi^{-1}$.  We seek to compare
$E = X_{r+1}Y_rX_r \cdots Y_1X_1$ with $X = X_{r+1} \cdots X_1$.
Again because products are associative, we may write $E = X_{r+1}((Y_rX_r) \cdots (Y_1X_1))$,
and each of the inner convolutions $e_i = y_i * x_i$ trivially collapses to uniform on $H$.  Further
the sequence of convolutions $v=e_r * \cdots * e_1$ remains uniform on $H$ and the final $x_{r+1} *v$ 
is uniform on $\pi H$.  On the other hand $x_2 * x_1$ has support on translate of double coset $H\pi H$
and continued left convolution can only make this count go up.  Clearly then, we have.
\begin{prop}
$X$ is more secure than $E$.
\end{prop}
\end{note}

\begin{note}{A Resource-Bounded Example of Extreme Expansion}
It may seem from our treatment in the above examples that the ciphers here are purely information
theoretic, applying only to infeasible and hypothetical ciphers.  In this section, we present a 
positive example of a computationally efficient alternating product cipher $T=XYZ$ which has nearly
optimal expansion of permutations along a huge double coset, yet where the $D=XY$ is trivially 
distinguishable from any idealized cipher.

To facilitate such a comparison, we exploit special properties of a polynomial-time cipher which 
achieves \emph{every} permutation (given enough key construction data or equivalently a private 
random function oracle).  

This construction from \cite{cis} called a \emph{universal security amplifier} was originally put 
forth to decided whether any efficient block cipher 
possessed a property of the one-time pad, namely that when composed with a non-perfect cipher was 
strictly more secure.

For security parameter $n$, let $\msg'=\{0,\ldots,2^n\}$ and let $X$ and $Z$ independent universal 
security amplifiers on $\msg=\{0,\ldots,2^n-1\}$, so they leave fixed the final plaintext $2^n$. Let 
$Y$ deterministically pick out any permutation $\pi\not\in\sym{\msg}$.  For example $\pi$ could 
simply add $1 \mod 2^n+1$, which is clearly computationally feasible.  Now we'd like to 
compare $T=XYZ$ with $D=XZ$.  
Since $X$ and $Z$ clearly have the group property, $\supp(d)=\sym{\msg}$, and so encrypting
once the plaintext $2^n$ will yield ciphertext $2^n$ with 100\% probability.  The following 
is thus immediate.

\begin{prop}
The product $D$ is distinguishable from any idealized cipher.
\end{prop}

We may further exploit Thm.~\ref{thm:3conv2cos} and in this case, the expansion is huge.

\begin{prop}
The alternating product cipher $T$ has $\supp(t)$ on about $(2^n+1)!$ permutations.
\end{prop}

\begin{pf} 
Because the action $\sym{\msg'}$ on $\msg'$ is multiply transitive we have by a standard result from 
group theory (see \cite[Thm.\ 9.6]{rotm}) that $\sym{\msg'} = \sym{\msg} \cup \sym{\msg}\pi\sym{\msg}$,
in other words the double coset relevant to Thm.~\ref{thm:3conv2cos}, $\sym{\msg}\pi\sym{\msg}$ is 
nearly the whole of $\sym{\msg'}$.  This double coset has size $(2^n+1)!-2^n!\approx(2^n+1)!$.

It remains then to show $t$ has full support on the double coset.  But $z$ has \emph{full} support
on $\sym{\msg}$ \cite{cis}, and by Thm.~\ref{thm:3conv2cos} each $z_i\majby z$ so it cannot have fewer 
permutations on each of the left cosets $\lambda_i \sym{\msg}$. This forces $\supp(t) = 
\sym{\msg}\pi\sym{\msg}$, and we are done.
\end{pf}
\end{note}

%
%
\bibliographystyle{plain}
\bibliography{../refs}

\startappendix

\ifthenelse{\boolean{extabs}}{}{ 
\begin{note}{$\alpha$-Guesswork}
\label{n:agw}
Bonneau has made an extensive study of the guessability of human-chosen secrets (see \cite{bonnphd}).  
In \cite[\S3.2.3]{bonnphd} an appealing security metric, \emph{$\alpha$-guesswork}, is introduced to 
model real-world guessing attacks:
$$
G_\alpha(X) = w_\alpha(X) - w_\alpha(X) \sum_{i=1}^{w_\alpha(X)}x_{[i]}
            + \sum_{i=1}^{w_\alpha(X)} i x_{[i]},
$$
where $w_\alpha(X)=\min\{i|\sum_{j=1}^i x_{[j]} \geq\alpha\}$ is the \emph{marginal guesswork}.  
As a function of $\alpha$, $w_\alpha(X)$ is piecewise constant 
$$
w_\alpha(X) =  i, \;\;\; \sum_{j=1}^{i-1} x_{[j]} < \alpha \leq \sum_{j=1}^i x_{[j]}
$$
Thus the following integral reduces to a sum of rectangle areas
$$
\int_0^{\alpha} w_\beta(X) d\beta 
  = \sum_{i=1}^{w_\alpha(X)} i \cdot \left( \sum_{j=1}^i x_{[j]} - \sum_{j=1}^{i-1} x_{[j]} \right)
  = \sum_{i=1}^{w_\alpha(X)} ix_{[i]},
$$
provided the final rectangle is whole, i.e.\ $\alpha=\sum_{j=1}^ix_{[j]}$ for some $i\in\ints$.  In
\cite{bonnphd}, this idea of rounding up to the nearest whole rectangle is given notation
$
\jbceil{\alpha} = \sum_{i=1}^{w_\alpha(X)} x_{[i]},
$
which we'll write $\jbceil{\alpha}_X$ when we need to emphasize the dependence on $X$.  With 
these observations, we may write $\alpha$-guesswork as
\begin{equation}
\label{eqn:agw}
G_\alpha(X) = (1-\jbceil{\alpha}) w_\alpha(X) + \int_0^{\jbceil{\alpha}} w_\beta(X) d\beta,
\end{equation}
which nicely captures Lorenz curve aspects of guessing.\footnote{
Just as the Gini coefficient and guesswork are related to area under some Lorenz curve (see 
\cite{indo}), here we find $\alpha$-guesswork to be the sum of two areas: the integral in 
(\ref{eqn:agw}) is the truncated area under the curve while $(1-\jbceil{\alpha}) w_\alpha(X)$ is 
the rectangular area to the right of the integral.}  Indeed by splitting the final rectangle of the 
integral, we may write for any $\hat{\alpha}$ situated $\alpha \leq \hat{\alpha} \leq 
\jbceil{\alpha}_X$,
\begin{equation}
\label{eqn:splitagw}
G_\alpha(X) = (1-\hat{\alpha}) w_\alpha(X) + \int_0^{\hat{\alpha}} w_\beta(X) d\beta.
\end{equation}
\begin{prop}
$\alpha$-guesswork is Schur-concave.
\end{prop}
\begin{pf}
Take $X\sim x$ and $Y\sim y$ with $x\majby y$ and first observe that $\jbceil{\alpha}_X$ is 
Schur-concave because
$$
\jbceil{\alpha}_X 
  = \sum_{i=1}^{w_\alpha(X)} x_{[i]}
  \geq \sum_{i=1}^{w_\alpha(Y)} x_{[i]}
  \geq \sum_{i=1}^{w_\alpha(Y)} y_{[i]}
  = \jbceil{\alpha}_Y.
$$
Thus $\alpha \leq \jbceil{\alpha}_Y \leq \jbceil{\alpha}_X$, and we use (\ref{eqn:splitagw})
to compare 
\begin{eqnarray*}
G_\alpha(X) &=& (1-\jbceil{\alpha}_Y) w_\alpha(X) + \int_0^{\jbceil{\alpha}_Y} w_\beta(X) d\beta \\
  &\geq& (1-\jbceil{\alpha}_Y) w_\alpha(Y) + \int_0^{\jbceil{\alpha}_Y} w_\beta(Y) d\beta \\
  &=& G_\alpha(Y),
\end{eqnarray*}
where we have used the Schur-concavity of $w_\alpha$.
\end{pf}
\begin{rem}
Of course, two distinct distributions might agree up to $w_\alpha(X)=w_\alpha(Y)$, so 
$\alpha$-guesswork is not strictly Schur-concave.
\end{rem}
\end{note}

} 

\begin{note}{Variation Distance to Uniformity}
The following lemma will prove quite useful.
\begin{lem}
\label{lem:varschur}
The variation distance to uniformity is Schur-convex.
\end{lem}

\ifthenelse{\boolean{extabs}}{
\begin{pf}
Suppose $x\majby y$ with $x,y\in\rpn$.  As a consequence of the definition,
$$
\var{u-x} = \sum_{i=1}^{k_x}x_{[i]} - \frac{k_x}{n},
\;\;\mathrm{and}\;\;
\var{u-y} = \sum_{i=1}^{k_y}y_{[i]} - \frac{k_y}{n},
$$
where $k_x=\max\setst{i}{x_{[i]}\geq 1/n}$ and $k_y=\max\setst{i}{y_{[i]}\geq 1/n}$.  If $k_x<k_y$, 
then
\begin{eqnarray*}
\sum_{i=1}^{k_y} y_{[i]} - \frac{k_y}{n} 
  &=& \sum_{i=1}^{k_x} y_{[i]} + \sum_{i=k_x+1}^{k_y} y_{[i]} 
      - \left( \frac{k_x}{n} + \frac{k_y-k_x}{n} \right) \\
  &=& \sum_{i=1}^{k_x} y_{[i]}  - \frac{k_x}{n}
      + \sum_{i=k_x+1}^{k_y} \left(  y_{[i]} - \frac{1}{n}\right) 
  \geq \sum_{i=1}^{k_x} x_{[i]}  - \frac{k_x}{n}.
\end{eqnarray*}
I.e., $\var{u-y} \geq \var{u-x}$.  In case $k_x\geq k_y$ the result follows \emph{mutatis mutandis}.
\end{pf}
}{
\begin{pf}
Suppose $x\majby y$ with $x,y\in\rpn$.  As a consequence of the definition,
$$
\var{u-x} = \sum_{i=1}^{k_x}x_{[i]} - \frac{k_x}{n}, 
\;\;\;\mathrm{and}\;\;\;
\var{u-y} = \sum_{i=1}^{k_y}y_{[i]} - \frac{k_y}{n},
$$
where $k_x=\max\setst{i}{x_{[i]}\geq 1/n}$ and $k_y=\max\setst{i}{y_{[i]}\geq 1/n}$.  
If $k_x=k_y$, then
\begin{eqnarray*}
\var{u-y} &=& \sum_{i=1}^{k_y} y_{[i]} - \frac{k_y}{n} \\
  &\geq& \sum_{i=1}^{k_x} x_{[i]} - \frac{k_x}{n} \\
  &=& \var{u-x}.
\end{eqnarray*}
If $k_x<k_y$, then
\begin{eqnarray*}
\var{u-y} &=& \sum_{i=1}^{k_y} y_{[i]} - \frac{k_y}{n} \\
  &=& \sum_{i=1}^{k_x} y_{[i]} + \sum_{i=k_x+1}^{k_y} y_{[i]} 
      - \left( \frac{k_x}{n} + \frac{k_y-k_x}{n} \right) \\
  &=& \sum_{i=1}^{k_x} y_{[i]}  - \frac{k_x}{n}
      + \sum_{i=k_x+1}^{k_y} \left(  y_{[i]} - \frac{1}{n}\right) \\
  &\geq& \sum_{i=1}^{k_x} x_{[i]}  - \frac{k_x}{n} \;\;=\;\; \var{u-x}.
\end{eqnarray*}
If however $k_x>k_y$, then we may write \cite[\S 1.A.1]{moa}
$$
\var{u-x} = \frac{q_x}{n} - \sum_{i=1}^{q_x} x_{(i)}, 
\;\;\;\mathrm{and}\;\;\;
\var{u-y} = \frac{q_y}{n} - \sum_{i=1}^{q_y} y_{(i)}, 
$$
where $q_x=\max\setst{i}{x_{(i)}\geq 1/n}$ and $q_y=\max\setst{i}{y_{(i)}\geq 1/n}$.  
Notice that $k_x>k_y$ implies $q_x<q_y$, so we have
\begin{eqnarray*}
\var{u-y} &=& \frac{q_y}{n} - \sum_{i=1}^{q_y} y_{(i)} \\
  &=& \left( \frac{q_x}{n} + \frac{q_y-q_x}{n} \right) 
      - \sum_{i=1}^{q_x} y_{(i)} - \sum_{i=q_x+1}^{q_y} y_{(i)} \\
  &=& \frac{q_x}{n} - \sum_{i=1}^{q_x} y_{(i)}  
      + \sum_{i=q_x+1}^{q_y} \left( \frac{1}{n} - y_{(i)} \right) \\
  &\geq& \frac{q_x}{n} - \sum_{i=1}^{q_x} x_{(i)} \;\;=\;\; \var{u-x}.
\end{eqnarray*}
\end{pf}
}
\end{note}

\ifthenelse{\boolean{extabs}}{ 
\begin{note}{Sketch of Proof of Thm.~\ref{thm:qschur}}
\begin{skof}{Thm.~\ref{thm:qschur}}
For arbitrary $q$-tuple $p\in\msgq$, let $H=\stab_{G}(p)$, and let $\{g_i\}$ be a left transversal 
for $H$ in $G$.  Then the two cases correspond to two 
different \emph{induced representations} from $H$ to $G$.  Specifically, $\reals\ind{H}{G}$ describes
distributions over $q$-tuples for comparing values of $\ncpa{X}(q)$, while $\reals H\ind{H}{G}$ 
describes the distributions over all permutations for comparing values of $W(X|C,p)$.

In either case, a general result adapted from \cite[Lem.~3.3]{cis} is that if $z\majby y$ in 
$V\ind{H}{G}\iso\bigoplus_i g_i\tens V$ and the permutations for this majorization in 
(\ref{eqn:hlpbvn}) act on it by permuting direct summands $g_i\tens V$ then 
\begin{equation}
\label{eqn:block}
z_\desc^{(i)} = \sum_{i=1}^{\idx{G}{H}} \omega_{ij} D_{ij} y_\desc^{(i)},
\end{equation}
where $z^{(i)}$ and $y^{(i)}$ are projections of $z$ and $y$ into the direct summands of $z$ and $y$, 
and where each $D_{ij}$ is doubly stochastic as is the matrix $\Omega=[\omega_{ij}]$. 

(\emph{Case 1:}) Define $\sigma$ taking $x\mapsto\hat{x}=\sum_i x_\desc^{(i)}$. The product
relation assures (\ref{eqn:block}), which by a result of Day \cite[Prop.~5.A.6]{moa}
implies $\hat{z}\majby\hat{y}$.  Since guesswork is Schur-concave and 
$W(X|C,p)=W(\sum_i x_\desc^{(i)})$ we have $z\qmajby y$.

(\emph{Case 2:}) Define $\sigma$ taking $x\mapsto\hat{x}=\transp{[x(g_1H),\ldots,
x(g_{\idx{G}{H}} H)]}$.  Likely beginning with \cite{dcor} and more recently \cite{mrs} 
NCPA advantage $\ncpa{X}(q)$ is identified with variation distance to uniformity 
$\var{\hat{x}-\hat{u}}$, which by Lem.~\ref{lem:varschur} is Schur-convex.  
Now the action of $G$ on $G/H$ also gives rise to a left module action of $\reals G$ on 
$\reals\ind{H}{G}$ consistent with (\ref{eqn:block}), now with 1-dimensional summands.  
The $1\times 1$ doubly stochastic matrices vanish and we obtain $\hat{z}=\Omega\hat{y}$ or 
$\hat{z}\majby\hat{y}$. Again $z\qmajby y$ by the Schur-convexity of $\ncpa{X}(q)$.
\end{skof}
\end{note}
}{ 
\input{fullpf.tex}
}
\end{document}

%% file: macros.tex
\newcommand{\hlp}{Hardy-Littlewood-P\'{o}lya}
\newcommand{\bvn}{Birkhoff-von Neumann}

\newcommand{\iso}{\cong}
\newcommand{\defeq}{\triangleq}
\newcommand{\map}{\rightarrow}

\newcommand{\tens}{\otimes}

\newcommand{\idx}[2]{[#1\mbox{:}#2]}

\newcommand{\setst}[2]{ \left\{ #1 \mid #2 \right\} }

\newcommand{\jbceil}[1]{\lceil\mkern-5mu\lceil#1\rceil\mkern-5mu\rceil}

\newcommand{\supp}{\mathrm{supp}}
\newcommand{\stab}{\mathrm{Stab}}
\newcommand{\lconj}[2]{{}^{#1}#2}

\newcommand{\transp}[1]{{#1}^\mathrm{t}}

\newcommand{\ind}[2]{\mbox{$\uparrow$}_{#1}^{#2}}

\newcommand{\prob}[1]{\mathsf{Pr}\left[#1\right]}

\newcommand{\var}[1]{\left\|#1\right\|}

\newcommand{\majby}{\preceq}

\newcommand{\smajby}{\prec}
\newcommand{\qmajby}{\preceq_q}
\newcommand{\desc}{\downarrow}

\newcommand{\ints}{\mathbb{Z}}

\newcommand{\reals}{\mathbb{R}}
\newcommand{\rpn}{\reals_+^n}

\newcommand{\msg}{\mathcal{M}}
\newcommand{\msgq}{\mathcal{M}^{(q)}}

\newcommand{\sym}[1]{\mathfrak{S}_{#1}}

\newcommand{\adv}{\mathbf{Adv}}
\newcommand{\ncpa}[1]{\adv_{#1}^\mathrm{ncpa}}

%% file: lncsnotes.tex
%
%

\renewenvironment{note}[1]{\subsection{#1.}}{}
\newcommand{\notesect}[1]{\section{#1}}

\newtheorem{thm}{Theorem}
\newtheorem{defn}{Definition}
\newtheorem{lem}{Lemma}
\newtheorem{prop}{Proposition}
\newtheorem{coro}{Corollary}
\newtheorem{ex}{Example}

\newenvironment{pf}{\paragraph*{}\hspace{-0.6em}\textit{Proof:}}{\hfill $\Box$}

\newenvironment{skof}[1]{\paragraph*{}\hspace{-0.6em}\textit{Proof Sketch of #1:}}{\hfill $\Box$}
\newenvironment{rem}{\paragraph*{}\hspace{-0.6em}\textsc{Remark.}}{\hfill $\Box$}

\pagestyle{plain}
\raggedbottom

\newcommand{\startappendix}{
\setcounter{secnumdepth}{4}
\setcounter{section}{0}
\setcounter{subsection}{0}
\renewcommand*{\thesubsection}{\Alph{section}.\arabic{subsection}}
\renewcommand*{\thesection}{Appendices}

\section{}
}

%% file: double.tex
\newcommand{\double}[2]{
\vcenter{\btexdraw
\drawdim mm
\linewd 0.300000
\move (4.000000 0.000000)
\lvec (11.000000 0.000000)
\move (11.000000 0.000000)
\lvec (9.000000 1.000000)
\lvec (10.000000 0.000000)
\lvec (9.000000 -1.000000)
\lvec (11.000000 0.000000)
\lfill f:0
\move (15.000000 0.000000)
\textref h:C v:C \htext{$ #1 $}
\move (19.000000 2.469136)
\rlvec (0.000000 -4.938272)
\rlvec (-8.000000 0.000000)
\rlvec (0.000000 4.938272)
\rlvec (8.000000 0.000000)
\move (19.000000 0.000000)
\lvec (26.000000 0.000000)
\move (26.000000 0.000000)
\lvec (24.000000 1.000000)
\lvec (25.000000 0.000000)
\lvec (24.000000 -1.000000)
\lvec (26.000000 0.000000)
\lfill f:0
\move (30.000000 0.000000)
\textref h:C v:C \htext{$ #2 $}
\move (34.000000 2.469136)
\rlvec (0.000000 -4.938272)
\rlvec (-8.000000 0.000000)
\rlvec (0.000000 4.938272)
\rlvec (8.000000 0.000000)
\move (34.000000 0.000000)
\lvec (41.000000 0.000000)
\move (41.000000 0.000000)
\lvec (39.000000 1.000000)
\lvec (40.000000 0.000000)
\lvec (39.000000 -1.000000)
\lvec (41.000000 0.000000)
\lfill f:0
\etexdraw}
}

%% file: triple.tex
\newcommand{\triple}[3]{
\vcenter{\btexdraw
\drawdim mm
\linewd 0.300000
\move (4.000000 0.000000)
\lvec (11.000000 0.000000)
\move (11.000000 0.000000)
\lvec (9.000000 1.000000)
\lvec (10.000000 0.000000)
\lvec (9.000000 -1.000000)
\lvec (11.000000 0.000000)
\lfill f:0
\move (15.000000 0.000000)
\textref h:C v:C \htext{$ #1 $}
\move (19.000000 2.469136)
\rlvec (0.000000 -4.938272)
\rlvec (-8.000000 0.000000)
\rlvec (0.000000 4.938272)
\rlvec (8.000000 0.000000)
\move (19.000000 0.000000)
\lvec (26.000000 0.000000)
\move (26.000000 0.000000)
\lvec (24.000000 1.000000)
\lvec (25.000000 0.000000)
\lvec (24.000000 -1.000000)
\lvec (26.000000 0.000000)
\lfill f:0
\move (30.000000 0.000000)
\textref h:C v:C \htext{$ #2 $}
\move (34.000000 2.469136)
\rlvec (0.000000 -4.938272)
\rlvec (-8.000000 0.000000)
\rlvec (0.000000 4.938272)
\rlvec (8.000000 0.000000)
\move (34.000000 0.000000)
\lvec (41.000000 0.000000)
\move (41.000000 0.000000)
\lvec (39.000000 1.000000)
\lvec (40.000000 0.000000)
\lvec (39.000000 -1.000000)
\lvec (41.000000 0.000000)
\lfill f:0
\move (45.000000 0.000000)
\textref h:C v:C \htext{$ #3 $}
\move (49.000000 2.469136)
\rlvec (0.000000 -4.938272)
\rlvec (-8.000000 0.000000)
\rlvec (0.000000 4.938272)
\rlvec (8.000000 0.000000)
\move (49.000000 0.000000)
\lvec (56.000000 0.000000)
\move (56.000000 0.000000)
\lvec (54.000000 1.000000)
\lvec (55.000000 0.000000)
\lvec (54.000000 -1.000000)
\lvec (56.000000 0.000000)
\lfill f:0
\etexdraw}
}

%% file: altprod.tex
%
%
%
%
%
%
%
%
%
%
%
%
%
%
%
%
%
%
%
%
%
%
%
%
%
%
%
%
%
%
%
%
%
%
%
%
%
%
%
%
%
%
%
%
%
%
%
%
%
%
%
%
%
%
%
%
%
%
%
%
%
%
%
%
%
%
%
%
%
%
%
%
%
%
%
%
%
%
%
%
%
%
%
%
%
%
%
%
%
%
%
%
%
%
%
%
%
%
%
%
%
%
%
%
%
%
%
%
%
%
%
%
%
%
%
%
%
%
%
%
%
%
%
%
%
%
%
%
%
%
\vcenter{\btexdraw
\drawdim mm
\linewd 0.300000
\move (0.000000 0.000000)
\textref h:C v:C \htext{$ X= $}
\move (4.000000 0.000000)
\lvec (12.000000 0.000000)
\move (12.000000 0.000000)
\lvec (10.000000 1.000000)
\lvec (11.000000 0.000000)
\lvec (10.000000 -1.000000)
\lvec (12.000000 0.000000)
\lfill f:0
\move (0.000000 -11.200000)
\textref h:C v:C \htext{$ E= $}
\move (4.000000 -11.200000)
\lvec (12.000000 -11.200000)
\move (12.000000 -11.200000)
\lvec (10.000000 -10.200000)
\lvec (11.000000 -11.200000)
\lvec (10.000000 -12.200000)
\lvec (12.000000 -11.200000)
\lfill f:0
\move (0.000000 -22.400000)
\textref h:C v:C \htext{$ Y= $}
\move (4.000000 -22.400000)
\lvec (28.000000 -22.400000)
\move (28.000000 -22.400000)
\lvec (26.000000 -21.400000)
\lvec (27.000000 -22.400000)
\lvec (26.000000 -23.400000)
\lvec (28.000000 -22.400000)
\lfill f:0
\move (80.000000 0.000000)
\textref h:C v:C \htext{$ \cdots $}
\move (84.000000 0.000000)
\lvec (92.000000 0.000000)
\move (92.000000 0.000000)
\lvec (90.000000 1.000000)
\lvec (91.000000 0.000000)
\lvec (90.000000 -1.000000)
\lvec (92.000000 0.000000)
\lfill f:0
\move (80.000000 -11.200000)
\textref h:C v:C \htext{$ \cdots $}
\move (84.000000 -11.200000)
\lvec (92.000000 -11.200000)
\move (92.000000 -11.200000)
\lvec (90.000000 -10.200000)
\lvec (91.000000 -11.200000)
\lvec (90.000000 -12.200000)
\lvec (92.000000 -11.200000)
\lfill f:0
\move (80.000000 -22.400000)
\textref h:C v:C \htext{$ \cdots $}
\move (84.000000 -22.400000)
\lvec (108.000000 -22.400000)
\move (108.000000 -22.400000)
\lvec (106.000000 -21.400000)
\lvec (107.000000 -22.400000)
\lvec (106.000000 -23.400000)
\lvec (108.000000 -22.400000)
\lfill f:0
\move (96.000000 0.000000)
\textref h:C v:C \htext{$ X_{r+1} $}
\move (100.000000 2.469136)
\rlvec (0.000000 -4.938272)
\rlvec (-8.000000 0.000000)
\rlvec (0.000000 4.938272)
\rlvec (8.000000 0.000000)
\move (100.000000 0.000000)
\lvec (108.000000 0.000000)
\move (108.000000 0.000000)
\lvec (106.000000 1.000000)
\lvec (107.000000 0.000000)
\lvec (106.000000 -1.000000)
\lvec (108.000000 0.000000)
\lfill f:0
\move (96.000000 -5.600000)
\textref h:C v:C \htext{$ \downarrow $}
\move (96.000000 -11.200000)
\textref h:C v:C \htext{$ X_{r+1} $}
\move (100.000000 -8.730864)
\rlvec (0.000000 -4.938272)
\rlvec (-8.000000 0.000000)
\rlvec (0.000000 4.938272)
\rlvec (8.000000 0.000000)
\move (100.000000 -11.200000)
\lvec (108.000000 -11.200000)
\move (108.000000 -11.200000)
\lvec (106.000000 -10.200000)
\lvec (107.000000 -11.200000)
\lvec (106.000000 -12.200000)
\lvec (108.000000 -11.200000)
\lfill f:0
\move (16.000000 0.000000)
\textref h:C v:C \htext{$ X_1 $}
\move (20.000000 2.469136)
\rlvec (0.000000 -4.938272)
\rlvec (-8.000000 0.000000)
\rlvec (0.000000 4.938272)
\rlvec (8.000000 0.000000)
\move (20.000000 0.000000)
\lvec (44.000000 0.000000)
\move (44.000000 0.000000)
\lvec (42.000000 1.000000)
\lvec (43.000000 0.000000)
\lvec (42.000000 -1.000000)
\lvec (44.000000 0.000000)
\lfill f:0
\move (16.000000 -5.600000)
\textref h:C v:C \htext{$ \downarrow $}
\move (16.000000 -11.200000)
\textref h:C v:C \htext{$ X_1 $}
\move (20.000000 -8.730864)
\rlvec (0.000000 -4.938272)
\rlvec (-8.000000 0.000000)
\rlvec (0.000000 4.938272)
\rlvec (8.000000 0.000000)
\move (20.000000 -11.200000)
\lvec (28.000000 -11.200000)
\move (28.000000 -11.200000)
\lvec (26.000000 -10.200000)
\lvec (27.000000 -11.200000)
\lvec (26.000000 -12.200000)
\lvec (28.000000 -11.200000)
\lfill f:0
\move (32.000000 -11.200000)
\textref h:C v:C \htext{$ Y_1 $}
\move (36.000000 -8.730864)
\rlvec (0.000000 -4.938272)
\rlvec (-8.000000 0.000000)
\rlvec (0.000000 4.938272)
\rlvec (8.000000 0.000000)
\move (36.000000 -11.200000)
\lvec (44.000000 -11.200000)
\move (44.000000 -11.200000)
\lvec (42.000000 -10.200000)
\lvec (43.000000 -11.200000)
\lvec (42.000000 -12.200000)
\lvec (44.000000 -11.200000)
\lfill f:0
\move (32.000000 -16.800000)
\textref h:C v:C \htext{$ \uparrow $}
\move (32.000000 -22.400000)
\textref h:C v:C \htext{$ Y_1 $}
\move (36.000000 -19.930864)
\rlvec (0.000000 -4.938272)
\rlvec (-8.000000 0.000000)
\rlvec (0.000000 4.938272)
\rlvec (8.000000 0.000000)
\move (36.000000 -22.400000)
\lvec (60.000000 -22.400000)
\move (60.000000 -22.400000)
\lvec (58.000000 -21.400000)
\lvec (59.000000 -22.400000)
\lvec (58.000000 -23.400000)
\lvec (60.000000 -22.400000)
\lfill f:0
\move (48.000000 0.000000)
\textref h:C v:C \htext{$ X_2 $}
\move (52.000000 2.469136)
\rlvec (0.000000 -4.938272)
\rlvec (-8.000000 0.000000)
\rlvec (0.000000 4.938272)
\rlvec (8.000000 0.000000)
\move (52.000000 0.000000)
\lvec (76.000000 0.000000)
\move (76.000000 0.000000)
\lvec (74.000000 1.000000)
\lvec (75.000000 0.000000)
\lvec (74.000000 -1.000000)
\lvec (76.000000 0.000000)
\lfill f:0
\move (48.000000 -5.600000)
\textref h:C v:C \htext{$ \downarrow $}
\move (48.000000 -11.200000)
\textref h:C v:C \htext{$ X_2 $}
\move (52.000000 -8.730864)
\rlvec (0.000000 -4.938272)
\rlvec (-8.000000 0.000000)
\rlvec (0.000000 4.938272)
\rlvec (8.000000 0.000000)
\move (52.000000 -11.200000)
\lvec (60.000000 -11.200000)
\move (60.000000 -11.200000)
\lvec (58.000000 -10.200000)
\lvec (59.000000 -11.200000)
\lvec (58.000000 -12.200000)
\lvec (60.000000 -11.200000)
\lfill f:0
\move (64.000000 -11.200000)
\textref h:C v:C \htext{$ Y_2 $}
\move (68.000000 -8.730864)
\rlvec (0.000000 -4.938272)
\rlvec (-8.000000 0.000000)
\rlvec (0.000000 4.938272)
\rlvec (8.000000 0.000000)
\move (68.000000 -11.200000)
\lvec (76.000000 -11.200000)
\move (76.000000 -11.200000)
\lvec (74.000000 -10.200000)
\lvec (75.000000 -11.200000)
\lvec (74.000000 -12.200000)
\lvec (76.000000 -11.200000)
\lfill f:0
\move (64.000000 -16.800000)
\textref h:C v:C \htext{$ \uparrow $}
\move (64.000000 -22.400000)
\textref h:C v:C \htext{$ Y_2 $}
\move (68.000000 -19.930864)
\rlvec (0.000000 -4.938272)
\rlvec (-8.000000 0.000000)
\rlvec (0.000000 4.938272)
\rlvec (8.000000 0.000000)
\move (68.000000 -22.400000)
\lvec (76.000000 -22.400000)
\move (76.000000 -22.400000)
\lvec (74.000000 -21.400000)
\lvec (75.000000 -22.400000)
\lvec (74.000000 -23.400000)
\lvec (76.000000 -22.400000)
\lfill f:0
\etexdraw}